\documentclass[preprint,sort&compress,times]{elsarticle}

\usepackage{amsmath}
\usepackage{graphicx}
\usepackage{amssymb}
\usepackage{bm}
\usepackage{array}
\usepackage{hyperref}
\usepackage{slashed}
\usepackage{braket}
\usepackage{multirow}
\usepackage{bigstrut}
\usepackage[nooneline,footnotesize,center]{caption2}
\usepackage{color}

\journal{Physics Letter B}

\begin{document}

\begin{frontmatter}

\title {Quark to $\Lambda$-hyperon spin transfers in the current-fragmentation region}

\author[PKU]{Yujie Chi}

\author[PKU,CHEP]{Bo-Qiang Ma\corref{cor1}}
\ead{mabq@pku.edu.cn}

\cortext[cor1]{Corresponding author}

\address[PKU]{School of Physics and State Key Laboratory of Nuclear Physics and
Technology, Peking University, Beijing 100871,
China}
\address[CHEP]{Center for High Energy
Physics, Peking University, Beijing 100871,
China}

\begin{abstract}
We perform a study on the struck quark to the $\Lambda$-hyperon fragmentation processes
by taking into account the anti-quark fragmentations and intermediate decays from other hyperons.
We concentrate on how the longitudinally polarized quark fragments to the
longitudinally polarized $\Lambda$, how unpolarized quark and anti-quark fragment to the unpolarized
$\Lambda$, and how quark and anti-quark fragment to the $\Lambda$ through the intermediate
decay processes. We calculate the effective fragmentation functions
in the light-cone SU(6) quark-spectator-diquark model via the Gribov-Lipatov relation, with
the Melosh-Wigner rotation effect also included. The calculated results are
in reasonable agreement with the HERMES semi-inclusive $ep$ experimental data and the OPAL and ALEPH $e^{+}e^{-}$ annihilation experimental data.
\end{abstract}

\end{frontmatter}

In high energy physics, the current-fragmentation (CF) region in
lepton-hadron semi-inclusive deep inelastic scattering
(SIDIS) is sensitive
to the quark distributions and fragmentations.
%In this region, when the parton is struck out from the target particle and
%then fragments into a final hadron, both the parton distribution functions (PDFs) in the
%target particle and the fragmentation functions (FFs) to the final hadron can be extracted.
In this kind of processes, the colored parton
inside the target is struck with great momentum
and then quickly fragments into final hadrons. If we make cross section measurement of one of the final hadrons,
both the target parton distribution functions (PDFs) and the quark to final hadron fragmentation functions (FFs) can be extracted.
In the study of the proton spin substructure,
the $\Lambda$-hyperon among the produced hadrons is suggested to be studied~\cite{Artru:1990wq,Cortes:1991ja,Jaffe:1991ra,Ellis:1995fc,Lu:1995np,Ma:1998pd}.
This is mainly due to the facts that the $\Lambda$-hyperon has relatively large production cross section and
that its polarization is self-analyzing owing to its characteristic decay mode $\Lambda \rightarrow p \pi ^{-}$
with a large branching ratio of $64\%$.

In the naive quark model, the spin of the
$\Lambda$-hyperon is carried by the $s$ quark and the $u,d$ quarks inside the $\Lambda$ formulate a spin and isospin zero state. If this correlation
conserves in the current-fragmentation process, it is reasonable to speculate that the total spin
transfer from the $u,d$ quarks to the $\Lambda$-hyperon
is zero. However,
the data from the deep inelastic scattering experiment imply that the spin transfers are none zero from the struck $u,d$ quarks
to the produced $\Lambda$-hyperon.
If this property can be carried over to the quark distributions inside the $\Lambda$-hyperon,
it means that the quark distributions of the $\Lambda$-hyperon are more interesting than the naive quark model predicted.
Previous studies discussed this issue and made various PDF and FF predictions for the $\Lambda$-hyperon~\cite{Nzar:1995wb,Jaffe:1996wp,Kotzinian:1997vd,Kotzinian:1997dp,deFlorian:1997zj,deFlorian:1998ba,Boros:1998kc,
Liang:1998sn,Ma:1999gj,Ma:1999wp,Ma:2000uu,Anselmino:2000ga,Ma:2001rm,Yang:2001pr}. Experimental data also indicate that in the current-fragmentation region, the $\Lambda$-hyperon may be produced through the intermediate decay processes of other hyperons.
Besides, we know that in the small $x$ region
the sea quark distributions are dominating over the valence quark distributions inside the proton.
If this correlation keeps in the fragmentation process, the small $z$ region of the
$\Lambda$ production cross section should be sensitive to the probability of the anti-quark distribution in the target particle
and the probability of the anti-quark fragmentation into the $\Lambda$.
There have been relevant discussions concerning anti-quark fragmentations~\cite{Anselmino:2001ps,Ellis:2007ig} and intermediate decays\cite{Boros:2000ex}.

In this Letter, we provide a first study combining both the intermediate decay processes and the anti-quark fragmentation processes
in the $\Lambda$ fragmentation process.
From QCD factorization theorem, the high energy collision cross section
can be calculated by using the perturbation theory complemented with the soft QCD effects embedded in
quark distributions and fragmentation functions, which are process insensitive and universal. If we take the $eP\rightarrow e\Lambda X$ process to extract the fragmentation functions (FFs) using the factorization theorem,
the same FFs should be applicable to the $e^{+}e^{-}$ annihilation process.

For a general $eP\rightarrow eP_h X$ process, the differential scattering cross
section at the tree level can be effectively expanded as
\begin{eqnarray}\label{factorism}
&&\mathrm{d}\sigma=\frac{1}{4\ell{\cdot}P}\sum_{s_{e^{'}}}\sum_{X}
\int\frac{d^{3}\overrightarrow{P}_{X}}{(2\pi)^{3}2E_{X}}
(2\pi)^{4}\delta^{4}(P+\ell-P_{X}-P_{h}-\ell^{'})\nonumber\\
&&
~~~~~\times \bigg\{ \frac{e^{4}}{q^{4}}\left[ \overline{u}_{e^{'}}(\ell^{'},s_{e^{'}})
\gamma_{\mu}u_{e}(\ell,s_{e})\right]^{*}
\left[\overline{u}_{e^{'}}(\ell^{'},s_{e^{'}})
\gamma_{\nu}u_{e}(\ell,s_{e})\right] \nonumber\\
&&
~~~~~~\times\left<X,P_{h}S_{h}|J^{\mu}(0)|PS\right>^{*}\left<X,P_{h}S_{h}|J^{\nu}(0)|PS\right> \bigg\} \frac{d^{3}\ell^{'}}{(2\pi)^{3}2E^{'}}
\frac{d^{3}\overrightarrow{P}_{h}}{(2\pi)^{3}2E_{h}}\nonumber\\
&&~~~=\frac{1}{4\ell{\cdot}P}\frac{e^{4}}{Q^{4}}L_{\mu\nu}W^{\mu\nu}(2\pi)^{4}
\frac{d^{3}\ell^{'}}{(2\pi)^{3}2E^{'}}
\frac{d^{3}\overrightarrow{P}_{h}}{(2\pi)^{3}2E_{h}},
\end{eqnarray}
where $L_{\mu\nu}$ and $W^{\mu\nu}$
are the leptonic tensor and the hadronic tensor respectively.

Defining three Lorentz invariants
\begin{equation}\label{invariant}
x=\frac{Q^{2}}{2P{\cdot}q},~y=\frac{P{\cdot}q}{P{\cdot}\ell},
~z=\frac{P{\cdot}P_{h}}{P{\cdot}q},
\end{equation}
we rewrite the cross section ~(\ref{factorism}) as
\begin{equation}\label{cross}
\frac{\mathrm{d}\sigma}{\mathrm{d}x\mathrm{d}y
\mathrm{d}z\mathrm{d}^2\overrightarrow{P}_{h\bot}}=
\frac{\pi\alpha_{\mathrm{em}}^2}{2Q^4}\frac{y}{z}L_{\mu\nu}W^{\mu\nu},
\end{equation}
where $\overrightarrow{P}_{h\bot}$ is the transverse momentum of the
produced hadron.

Then in the tree level, the leptonic tensor can be decomposed into a symmetric and an antisymmetric part as
\begin{eqnarray}\label{leptonic}
&&L_{\mu\nu}=\sum_{s_{e^{'}}}\left[\overline{u}_{e^{'}}(\ell^{'},s_{e^{'}})
\gamma_{\mu}u_{e}(\ell,s_{e})\right]^{\ast}\left[\overline{u}_{e^{'}}(\ell^{'},s_{e^{'}})
\gamma_{\nu}u_{e}(\ell,s_{e})\right]
\nonumber\\
&&~~~=2(\ell_{\mu}\ell_{\nu}^{'}+\ell_{\nu}\ell_{\mu}^{'}-g_{\mu\nu}\ell{\cdot}
\ell^{'})+2\textrm{i}\lambda_{e}\varepsilon_{\mu\nu\rho\sigma}
\ell^{\rho}q^{\sigma}.
\end{eqnarray}

In the parton model, the hadronic tensor is a convolution of PDFs and FFs.
At the twist two level, if we use a polarized electron beam to hit an unpolarized proton target,
both the unpolarized quark fragmentation function and the helicity-dependent quark fragmentation function
can be extracted. That is
\begin{eqnarray}\label{hadronic}
&&W^{\mu\nu}=\frac{1}{(2\pi)^4}\sum_{X}\int\frac{d^{3}\overrightarrow{P}_{X}}
{(2\pi)^{3}2E_{X}}(2\pi)^{4}\delta^{4}(P+\ell-P_{X}-P_{h}-\ell^{'})\nonumber\\
&&~~~~~~\left<X,P_{h}S_{h}|J^{\mu}(0)|PS\right>^{*}\left<X,P_{h}S_{h}|J^{\nu}(0)|PS\right>\nonumber\\
&&~~~=\sum_{a}e_a^2\int\frac{\mathrm{d}k^{-}\mathrm{d}^2
\overrightarrow{k}_T}{(2\pi)^4}\int\frac{\mathrm{d}\kappa^{+}\mathrm{d}^2
\overrightarrow{\kappa}_T}{(2\pi)^4}
\delta^2(\overrightarrow{k}_T-\overrightarrow{q}_T-
\overrightarrow{\kappa}_T)\nonumber\\
&&~~~~~~{\mathrm{Tr}}\left[\frac{1}{2}f_a(x,\overrightarrow{k}_T^2)\slashed{P}
\gamma^{\mu}\frac{1}{2}(D_a(z,\overrightarrow{\kappa'}^{2}_T)
\slashed{P_h}+\lambda_h\Delta
D_a(z,\overrightarrow{\kappa'}^{2}_T)\gamma_5)
\slashed{P_h}\right]_{k^+=xP^+,\kappa^-=P^-_h/z},
\end{eqnarray}
where $f_a(x,\overrightarrow{k}_T^2)$
is the unpolarized quark distribution in the proton, and $D_a(z,\overrightarrow{\kappa'}^{2}_T)$ and $\Delta
D_a(z,\overrightarrow{\kappa'}^{2}_T)$ indicate the probability of an unpolarized quark fragments into an unpolarized hadron
and the probability of a longitudinally polarized quark into a longitudinally polarized
hadron respectively.

The helicity asymmetry cross section is then obtained as
\begin{eqnarray}\label{assymetry}
&&A(x,y,z)=\frac{\mathrm{d}\sigma_\|-\mathrm{d}\sigma_{-\|}}
{\mathrm{d}\sigma_\|+\mathrm{d}\sigma_{-\| }}\nonumber\\
&&~~~=\frac{\frac{4\pi\alpha_{\mathrm{em}}^2s}{Q^4}
\sum_{a}e_a^2xy(1-y/2)f_a(x,Q^2)\Delta D_a(z,Q^2)}
{\frac{4\pi\alpha_{\mathrm{em}}^2s}{Q^4}\sum_{a}e_a^2x\frac{1+(1-y)^2}{2}
f_a(x,Q^2)D_a(z,Q^2)}\nonumber\\
&&~~~=\frac{y(2-y)}{1+(1-y)^2}\frac{\sum_{a}e_a^2xf_a(x,Q^2)
\Delta D_a(z,Q^2)}{\sum_{a}e_a^2xf_a(x,Q^2)D_a(z,Q^2)}.
\end{eqnarray}

From Eq.~(\ref{assymetry}), we take the part
\begin{eqnarray}\label{spintransfer}
&&A(x,z)=\frac{\sum_{a}e_a^2xf_a(x,Q^2)
\Delta D_a(z,Q^2)}{\sum_{a}e_a^2xf_a(x,Q^2)D_a(z,Q^2)}
\end{eqnarray}
as the longitudinal spin transfer factor.
Previous work reduced the common factor $x$ in Eq.~(\ref{spintransfer}), which is reasonable
in present experimental region. However, if the experimental measurement can make bins with broad $x$, theoretical
calculation should make an integration over $x$. This means that the $x$ factor in Eq.~(\ref{spintransfer})
should not be neglected.

A Monte Carlo calculation using the LEPTO generator
indicates that only about 40\%-50\% of $\Lambda$'s are produced directly,
30\%-40\% originate from $\Sigma^*$(1385) decay and about 20\% are
decay products of the $\Sigma^{0}$. The COMPASS Collaboration
measured the relative weights of the $\Sigma^{\ast}$ and the $\Xi$-hyperon decaying to the $\Lambda \pi$.
The results are about $20\%$ smaller than the Monte Carlo calculation~\cite{Adolph:2013dhv}.

Effectively, we can rewrite the helicity-dependent fragmentation function
$\Delta D_a(z,Q^2)$ and the unpolarized fragmentation function $D_a(z,Q^2)$ of the $\Lambda$ as
\begin{eqnarray}\label{eq:po}
\Delta D^{\Lambda}_{q}(z,Q^2)&=&a_1\Delta D_{q}^{\Lambda_{\mathrm{(direct)}}}(z,Q^2)
+a_2\Delta D_{q}^{\Sigma^{0}}(z,Q^2)\alpha_{\Sigma^{0}\Lambda}\\ \nonumber%
&&~~+a_3\Delta D_{q}^{\Sigma^{\ast}}(z,Q^2)\alpha_{\Sigma^{\ast}\Lambda}
+a_4\Delta D_{q}^{\Xi}(z,Q^2)\alpha_{\Xi\Lambda},
\end{eqnarray}
and
\begin{eqnarray}\label{eq:unpo}
&&D^{\Lambda}_{q}(z,Q^2)=a_1D_{q}^{\Lambda_{\mathrm{(direct)}}}(z,Q^2)+a_2D_{q}^{\Sigma^{0}}(z,Q^2)+a_3D_{q}^{\Sigma^{\ast}}(z,Q^2)+
a_4D_{q}^{\Xi}(z,Q^2).
\end{eqnarray}
Here, the weight coefficients are adjusted as
\begin{eqnarray}\label{a}
a_1=0.4,~a_2=0.2,~a_3=0.3,~a_4=0.1,
\end{eqnarray}
based on the spirit of the Monte Carlo prediction.

In the specific calculation, the weight coefficients of the $\Sigma^{\ast}$ is divided by three types of particles,
that is $\Sigma^{+}(1385)$, $\Sigma^{0}(1385)$ and $\Sigma^{-}(1385)$.
So the contribution to the spin transfer from the $\Sigma^{\ast}$ is
actually a mixture of these three hyperon decays. To simplify
the issue, we take 10\% of each branch for an average. The same treatment is done to
the $\Xi$, which contains the contribution from the $\Xi^{0}$ and $\Xi^{-}$, and 5\% of each branch is taken into consideration.

The $\alpha$'s are decay parameters, representing the
polarization transfer from the decay hyperon to the
$\Lambda$. In our study, these parameters are set as
\begin{eqnarray}
\alpha_{\Sigma^{0}\Lambda}=-0.333,~\alpha_{\Sigma^{\ast}\Lambda}=0.6,
~\alpha_{\Xi^{0}\Lambda}=-0.406, ~\alpha_{\Xi^{-}\Lambda}=-0.458,
\end{eqnarray}
where $\alpha_{\Sigma^{0}\Lambda}$ is the decay parameter of the process $\Sigma^{0} \rightarrow \Lambda \gamma$
discussed in ref.~\cite{Gatto:1958}, $\alpha_{\Xi^{0}\Lambda}$ and $\alpha_{\Xi^{-}\Lambda}$ are decay
parameters measured in experiments and their specific values are taken from~\cite{Beringer:2012}, and
$\alpha_{\Sigma^{\ast}\Lambda}$ is an estimated parameter by us. The choice of an $\alpha_{\Sigma^{\ast}\Lambda} = 0.6$ is
due to the facts that the spin of $\Sigma^{\ast}$ (being 3/2) should be almost total positively correlated with $\Lambda$ spin (being 1/2)
in the decay process corresponding to the $(s,s_z)=(3/2,\pm 3/2)$ components of $\Sigma^{\ast}$ and that there should be a suppression for
the spin transfer corresponding to the $(s,s_z)=(3/2,\pm 1/2)$ components of $\Sigma^{\ast}$.

In the intermediate decay process, the longitudinal
momentum fraction of the $\Lambda$ to the splitting quark should be less than the longitudinal
fraction of the decay hyperon to the splitting quark.
In the light-cone formalism, the momentum fraction $z$ is defined as
$z=\frac{P_h^{-}}{q^{-}}$. This effect is taken into account by redefining
$\frac{P_{h}^{-}}{q^{-}}=1.1\ast\frac{P_{\Lambda}^-}{q^-}$.

In the year 1989, the polarized deeply inelastic scattering (DIS) experiment carried by
the European Muon Collaboration revealed that the sum of the helicity of the
quarks inside the proton is
much smaller than the spin of the proton~\cite{Ashman:1987hv,Ashman:1989ig}. This discovery is against the naive $SU(6)$ quark model prediction,
causing the so-called ``proton spin crisis" or ``proton spin puzzle".
One possible explanation to understand this puzzle~\cite{Ma:1991xq,Ma:1992sj} is to take into account the relativistic effect of the quark transversal motions,
i.e., the Melosh-Wigner rotation effect~\cite{Melosh:1974cu}.
Based on this spirit, one can construct the light-cone SU(6) quark-spectator-diquark model to calculate
the valence quark spin distributions in the light-cone formalism~\cite{Ma:1996np,Ma:1997gy}.

We can also consider the Melosh-Wigner rotation effect
in the fragmentation process, and apply the light-cone SU(6) quark-spectator-diquark model
to estimate the probability of a valence quark directly
fragmenting to a hadron. This correlation can be realized through the
phenomenology Gribov-Lipatov relation~\cite{Gribov:1971zn,Gribov:1972rt,Brodsky:1996cc,Barone:2000tx}
\begin{equation}\label{glrelation}
    D_{q}^{h}(z){\sim}zq_{h}(z),
\end{equation}
where the fragmentation function $D_{q}^{h}(z)$ indicates a quark $q$
splitting into a hadron $h$ with longitudinal momentum fraction $z$, and
the distribution function $q_{h}(z)$ presents the probability of finding
the same quark $q$ carrying longitudinal momentum fraction $z$ inside the same hadron $h$.

The main idea of the light-cone SU(6) quark-spectator-diquark model is to
start from the naive SU(6) wave function of the hadron and then if any one of the quarks is probed, to reorganize
the other two quarks in terms of two quark wave functions with
spins 0 or 1 (scalar and vector diquarks), i.e., the diquark
being served as an effective particle which is called the spectator.

The unpolarized quark distribution for a quark with flavor $q$
inside a hadron $h$ is expressed as
\begin{equation}
q(x)= c_q^S a_S(x) + c_q^V a_V(x),
\end{equation}
where $c_q^S$ and $c_q^V$ are the weight coefficients determined by the SU(6) wave function,
and $a_D(x)$ ($D=S$ for scalar spectator or $V$ for axial vector spectator) denotes the amplitude for quark $q$ to be scattered while the
spectator is in the diquark state $D$. When expressed in terms of the
light-cone momentum space wave function $\varphi_{D} (x, {\mathbf k}_\perp)$, $a_D(x)$ reads
\begin{equation}
a_{D}(x) \propto  \int\left[\rm{d}^2 {\mathbf k}_\perp\right] |\varphi_{D} (x,
{\mathbf k}_\perp)|^2, \hspace{1cm} (D=S \hspace{0.2cm} {\mathrm{or}}
\hspace{0.2cm} V),
\end{equation}
and the normalization satisfies $\int_0^1 {\mathrm d} x a_D(x)=3$. To obtain a
practical formalism of the $a_D(x)$, we employ the
Brodsky-Huang-Lepage (BHL) prescription~\cite{Brodsky:1981jv} of the
light-cone momentum space wave function
\begin{equation}
\varphi_{D} (x, {\mathbf k}_\perp) = A_D \exp \left\{-\frac{1}{8\alpha_D^2}
\left[\frac{m_q^2+{\mathbf k}_\perp ^2}{x} + \frac{m_D^2+{\mathbf
k}_\perp^2}{1-x}\right]\right\},
\end{equation}
with the parameter $\alpha_D=330$~MeV. We set $\alpha_S=\alpha_V$ for $\alpha_D$'s
in our discussion because the non-perturbative physical effects can be effectively
reflected in the scalar and vector diquark masses. More detailed study should consider the difference in $\alpha_D$'s between scalar and vector diquarks.
The parameter of
the quark mass $m_q$ is the constituent quark mass and the scalar (vector) diquark mass $m_{D}$
($D=S,V$) is just an estimation from the constituent quark masses and the baryon masses.
This parametrization can reduce the free parameters to only a few, which are listed in Table 1.

The polarized quark distributions are obtained by introducing the Melosh-Wigner correction factor
\cite{Ma:1991xq,Ma:1992sj,Melosh:1974cu}
\begin{equation}
\Delta q(x)= \tilde{c}_q^S \tilde{a}_S(x) + \tilde{c}_q^V
\tilde{a}_V(x),
\end{equation}
where the coefficients $\tilde{c}_q^S$ and $\tilde{c}_q^V$ are
also determined by the SU(6) quark-diquark wave function, and
$\tilde{a}_D(x)$ is expressed as
\begin{equation}
\tilde{a}_{D}(x) = \int \left[\rm{d}^2 {\mathbf k}_\perp\right]
W_D(x,{\mathbf k}_\perp) |\varphi_{D} (x, {\mathbf k}_\perp)|^2,
\hspace{1cm} (D=S \hspace{0.2cm} {\mathrm{or}} \hspace{0.2cm} V),
\end{equation}
where
\begin{equation}
W_D(x,{\mathbf k}_{\perp}) =\frac{(k^+
+m_q)^2-{\mathbf k}^2_{\perp}} {(k^+ +m_q)^2+{\mathbf
k}^2_{\perp}} \label{eqM1},
\end{equation}
with $k^+=x {\cal M}$ and ${\cal M}^2=\frac{m^2_q+{\mathbf
k}^2_{\perp}}{x}+\frac{m^2_D+{\mathbf k}^2_{\perp}}{1-x}$. The
weight coefficients are also listed in Table
1. In this model, though the mass difference between different quarks and diquarks breaks the SU(3) symmetry explicitly, the SU(3)
symmetry between the octet baryons is in principle maintained in formalism.

\vspace{0.5cm}
%\newpage

\begin{footnotesize}
\centerline{Table~1~~ The quark distribution functions of octet
baryons in the SU(6) quark-diquark model~\cite{Ma:2000cg}}

\vspace{0.3cm}

\begin{center}
\begin{tabular}{|c||c|c||c|c||c|c|c|}\hline
 ~~~~~~~~~~~~~~~& $~~~~~~~$ & ~~~~~~~~~~~~~~~& $~~~~~~$ &~~~~~~~~~~~~~~~&
$m_q$ & $m_V$ & $m_S$
\\
Baryon& $~~q ~~$ & ~~~~~~~~~~~~~~~& $~~\Delta q ~~$
&~~~~~~~~~~~~~~~& (MeV) & (MeV) & (MeV)
\\ \hline
 ~~~~p~~~~ & $~~u~~$ &$\frac{1}{6}a_V+\frac{1}{2}a_S $ &
 $\Delta u$ & -$\frac{1}{18}\tilde{a}_V+\frac{1}{2}\tilde{a}_S $ &
330 & 800 & 600 \\ \cline{2-8}
 (uud)& $~~d~~$ &$\frac{1}{3}a_V$ &
 $\Delta d$ & -$\frac{1}{9}\tilde{a}_V$ &
330 & 800 & 600 \\ \cline{1-8}
 ~~~~n~~~~ & $~~u~~$ &$\frac{1}{3}a_V $ &
 $\Delta u$ & -$\frac{1}{9}\tilde{a}_V $ &
330 & 800 & 600 \\ \cline{2-8}
 (udd)& $~~d~~$ &$\frac{1}{6}a_V+\frac{1}{2} a_S$ &
 $\Delta d$ & -$\frac{1}{18}\tilde{a}_V+\frac{1}{2}\tilde{a}_S$ &
330 & 800 & 600 \\ \cline{1-8}
 $~~~~\Sigma^{+}~~~~$ & $~~u~~$ &$\frac{1}{6}a_V+\frac{1}{2}a_S $ &
 $\Delta u$ & -$\frac{1}{18}\tilde{a}_V+\frac{1}{2}\tilde{a}_S $ &
330 & 950 & 750 \\ \cline{2-8}
 (uus)& $~~s~~$ &$\frac{1}{3}a_V$ &
 $\Delta s$ & -$\frac{1}{9}\tilde{a}_V$ &
480 & 800 & 600 \\ \cline{1-8}
 $~~~~\Sigma^{0}~~~~$ & $~~u~~$ &$\frac{1}{12}a_V+\frac{1}{4}a_S $ &
 $\Delta u$ & -$\frac{1}{36}\tilde{a}_V+\frac{1}{4}\tilde{a}_S $ &
330 & 950 & 750 \\ \cline{2-8}
 (uds)& $~~d~~$ & $\frac{1}{12}a_V+\frac{1}{4}a_S $ &
 $\Delta d$ & -$\frac{1}{36}\tilde{a}_V+\frac{1}{4}\tilde{a}_S $ &
330 & 950 & 750 \\ \cline{2-8}
 $~~~~$ & $~~s~~$ & $\frac{1}{3}a_V$ &
 $\Delta s$ & -$\frac{1}{9}\tilde{a}_V $ &
480 & 800 & 600 \\ \cline{1-8} $~~~~\Sigma^{-}~~~~$ & $~~d~~$
&$\frac{1}{6}a_V+\frac{1}{2}a_S $ &
 $\Delta d$ & -$\frac{1}{18}\tilde{a}_V+\frac{1}{2}\tilde{a}_S $ &
330 & 950 & 750 \\ \cline{2-8}
 (dds)& $~~s~~$ &$\frac{1}{3}a_V$ &
 $\Delta s$ & -$\frac{1}{9}\tilde{a}_V$ &
480 & 800 & 600 \\ \cline{1-8} $~~~~\Lambda^{0}~~~~$ & $~~u~~$
&$\frac{1}{4}a_V+\frac{1}{12}a_S $ &
 $\Delta u$ & -$\frac{1}{12}\tilde{a}_V+\frac{1}{12}\tilde{a}_S $ &
330 & 950 & 750 \\ \cline{2-8}
 (uds)& $~~d~~$ & $\frac{1}{4}a_V+\frac{1}{12}a_S $ &
 $\Delta d$ & -$\frac{1}{12}\tilde{a}_V+\frac{1}{12}\tilde{a}_S $ &
330 & 950 & 750 \\ \cline{2-8}
 $~~~~$ & $~~s~~$ & $\frac{1}{3}a_S$ &
 $\Delta s$ & $\frac{1}{3}\tilde{a}_S $ &
480 & 800 & 600 \\ \cline{1-8}
 $~~~~\Xi^{-}~~~~$ & $~~d~~$ &$\frac{1}{3}a_V $ &
 $\Delta d$ & -$\frac{1}{9}\tilde{a}_V $ &
330 & 1100 & 900 \\ \cline{2-8}
 (dss)& $~~s~~$ &$\frac{1}{6}a_V+\frac{1}{2} a_S$ &
 $\Delta s$ & -$\frac{1}{18}\tilde{a}_V+\frac{1}{2}\tilde{a}_S$ &
480 & 950 & 750 \\ \cline{1-8}
 $~~~~\Xi^{0}~~~~$ & $~~u~~$ &$\frac{1}{3}a_V $ &
 $\Delta u$ & -$\frac{1}{9}\tilde{a}_V $ &
330 & 1100 & 900 \\ \cline{2-8}
 (uss)& $~~s~~$ &$\frac{1}{6}a_V+\frac{1}{2} a_S$ &
 $\Delta s$ & -$\frac{1}{18}\tilde{a}_V+\frac{1}{2}\tilde{a}_S$ &
480 & 950 & 750 \\ \cline{1-8}
\end{tabular}
\end{center}
\end{footnotesize}

\vspace{0.5cm}

Based on the same spirit, we give the distribution functions for the
$\Sigma^{\ast}$-hyperon, which in the naive quark model is a member of the SU(3) decuplet
with the total spin of 3/2.
Here, we try to use the same parameters to estimate
both the helicity and quark distribution functions in the light-cone SU(6)
quark-spectator-diquark model based on the following reasons:
(1) the mass of $\Sigma^{\ast}$ (which is about 1385 MeV) is similar to that of
$\Xi^-$ (which is about 1321 MeV), so we can use the same effective quark mass
parameters; (2) the total quark orbital angular momentum of $\Sigma^{\ast}$ is $0$,
so to form a spin $3/2$ particle, the diquark can only be in the vector state.
The specific helicity-dependent and unpolarized quark distribution functions for
the $\Sigma^{\ast}$'s in the quark-spectator-diquark model are shown in Table 2.

\vspace{0.5cm}
%\newpage
\begin{footnotesize}
\centerline{Table~2~~ The quark distribution functions of $\Sigma(1385)$'s $(s,s_z)=(3/2,\pm 3/2)$ components
in the light-cone SU(6) quark-spectator-diquark model}

\vspace{0.3cm}

\begin{center}
\begin{tabular}{|c||c|c||c|c||c|c|}\hline
 ~~~~~~~~~~~~~~~& $~~~~~~~$ & ~~~~~~~~~~~~~~~& $~~~~~~$ &~~~~~~~~~~~~~~~&
$m_q$ & $m_V$
\\
Baryon& $~~q ~~$ & ~~~~~~~~~~~~~~~& $~~\Delta q ~~$
&~~~~~~~~~~~~~~~& (MeV) & (MeV)
\\ \hline
 $~~~~\Sigma^{+}(1385)~~~~$ & $~~u~~$ &$\frac{2}{3}a_V$ &
 $\Delta u$ & $\frac{2}{3}\tilde{a}_V$ &
330 & 950  \\ \cline{2-7}
 (uus)& $~~s~~$ &$\frac{1}{3}a_V$ &
 $\Delta s$ & $\frac{1}{3}\tilde{a}_V$ &
480 & 800  \\ \cline{1-7}
 $~~~~\Sigma^{0}(1385)~~~~$ & $~~u~~$ &$\frac{1}{3}a_V$ &
 $\Delta u$ & $\frac{1}{3}\tilde{a}_V$ &
330 & 950 \\ \cline{2-7}
 (uds)& $~~d~~$ & $\frac{1}{3}a_V$ &
 $\Delta d$ & $\frac{1}{3}\tilde{a}_V$ &
330 & 950 \\ \cline{2-7}
 $~~~~$ & $~~s~~$ & $\frac{1}{3}a_V$ &
 $\Delta s$ & $\frac{1}{3}\tilde{a}_V $ &
480 & 800 \\ \cline{1-7} $~~~~\Sigma^{-}(1385)~~~~$ & $~~d~~$
&$\frac{2}{3}a_V$ &
 $\Delta d$ & $\frac{2}{3}\tilde{a}_V$ &
330 & 950 \\ \cline{2-7}
 (dds)& $~~s~~$ &$\frac{1}{3}a_V$ &
 $\Delta s$ & $\frac{1}{3}\tilde{a}_V$ &
480 & 800  \\ \cline{1-7}

\end{tabular}
\end{center}
\end{footnotesize}

\vspace{0.5cm}

We know that in the naive quark model, there is a SU(3) flavor
symmetry relation between octet baryons. We consider the anti-quark distribution inside
the octet baryons in the same way. To compare with the experimental data, the CTEQ5
parametrization (ctq5l) for proton is used as an input:
\begin{eqnarray}\label{antiquark}
u^{p}_{v}(x) &=& u^{\mathrm{ctq}}_{v}(x),\\ \nonumber %
d^{\Lambda}_{v}(x) &=& u^{\Lambda}_{v}(x)=\frac{u^{\Lambda,\mathrm{th}}_{v}(x)}{u^{p,\mathrm{th}}_{v}(x)}\ast u^{\mathrm{ctq}}_{v}(x),\\ \nonumber %
s^{\Lambda}_{v}(x) &=& \frac{s^{\Lambda,\mathrm{th}}_{v}(x)}{u^{p,\mathrm{th}}_{v}(x)}\ast u^{\mathrm{ctq}}_{v}(x),\\ \nonumber %
\Delta d^{\Lambda}_{v}(x) &=& \Delta u^{\Lambda}_{v}(x) = \frac{\Delta u^{\Lambda,\mathrm{th}}_{v}(x)}{u^{p,\mathrm{th}}_{v}(x)}\ast u^{\mathrm{ctq}}_{v}(x),\\ \nonumber %
\Delta s^{\Lambda}_{v}(x) &=& \frac{\Delta s^{\Lambda,\mathrm{th}}_{v}(x)}{u^{p,\mathrm{th}}_{v}(x)}\ast u^{\mathrm{ctq}}_{v}(x),\\ \nonumber %
d^{\Lambda}_{s}(x)&=&u^{\Lambda}_{s}(x) = \bar{u}^{\Lambda}(x)=\frac{1}{2}(\bar{u}^{\mathrm{ctq}}(x)+\bar{s}^{\mathrm{ctq}}(x)),\\ \nonumber %
s^{\Lambda}_{s}(x)&=&\bar{s}^{\Lambda}(x)=\bar{d}^{\mathrm{ctq}}(x),\nonumber %
\end{eqnarray}
where the $u^{\mathrm{ctq}}_{v}(x)$ means the PDF for the valence $u$ quark inside the proton from the CTEQ5 parametrization,
and the $u^{\Lambda,\mathrm{th}}_{v}(x)$ is the
PDF for the valence $u$ quark inside the $\Lambda$ given by the light-cone SU(6) quark-diquark model, so as other flavors.
For the other hyperons, the same spirit is followed. Apply the Gribov-Lipatov relation again, we can obtain the anti-quark FFs to the same hyperon.

Using all these equations from ~(\ref{a}) to~(\ref{glrelation}),~(\ref{antiquark}), and Tables 1 and 2, we can obtain
the effective $\Lambda$ fragmentation functions expressed in Eqs.~(\ref{eq:po}) and~(\ref{eq:unpo}), and the results are shown in Fig.~1.
The ratios of $\Delta D^u/D^u$ and $\Delta D^s/D^s$ with contributions from different channels are also plotted in Fig.~\ref{ratios}.
It is interesting that in the $z$ region we calculated, both $u$ and $s$ quarks contribute a positive spin transfer
(where the FFs of the $d$ quark is the same as that of the $u$ quark), when the direct fragmentation process and the intermediate decay process
are all considered. The anti-quark contribute a large FFs at the small $z$ region, as predicted. If we start from the struck quark, and end with the
$\Lambda$-hyperon, the FFs obtained from our method can be taken as an effective input to extract the PDFs of the target particle.

\begin{figure}%[htb]
\centering
\includegraphics[width=4in]{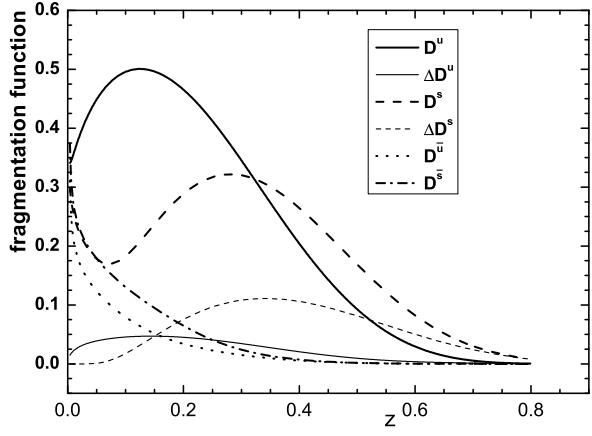}
\caption[*]{\baselineskip 13pt The results of the
$z$-dependent quark fragmentation functions to the $\Lambda$-hyperon, including unpolarized FFs and
helicity-dependent FFs.}
\end{figure}

\begin{figure}%[htb]
\centering
\includegraphics[width=2.35in]{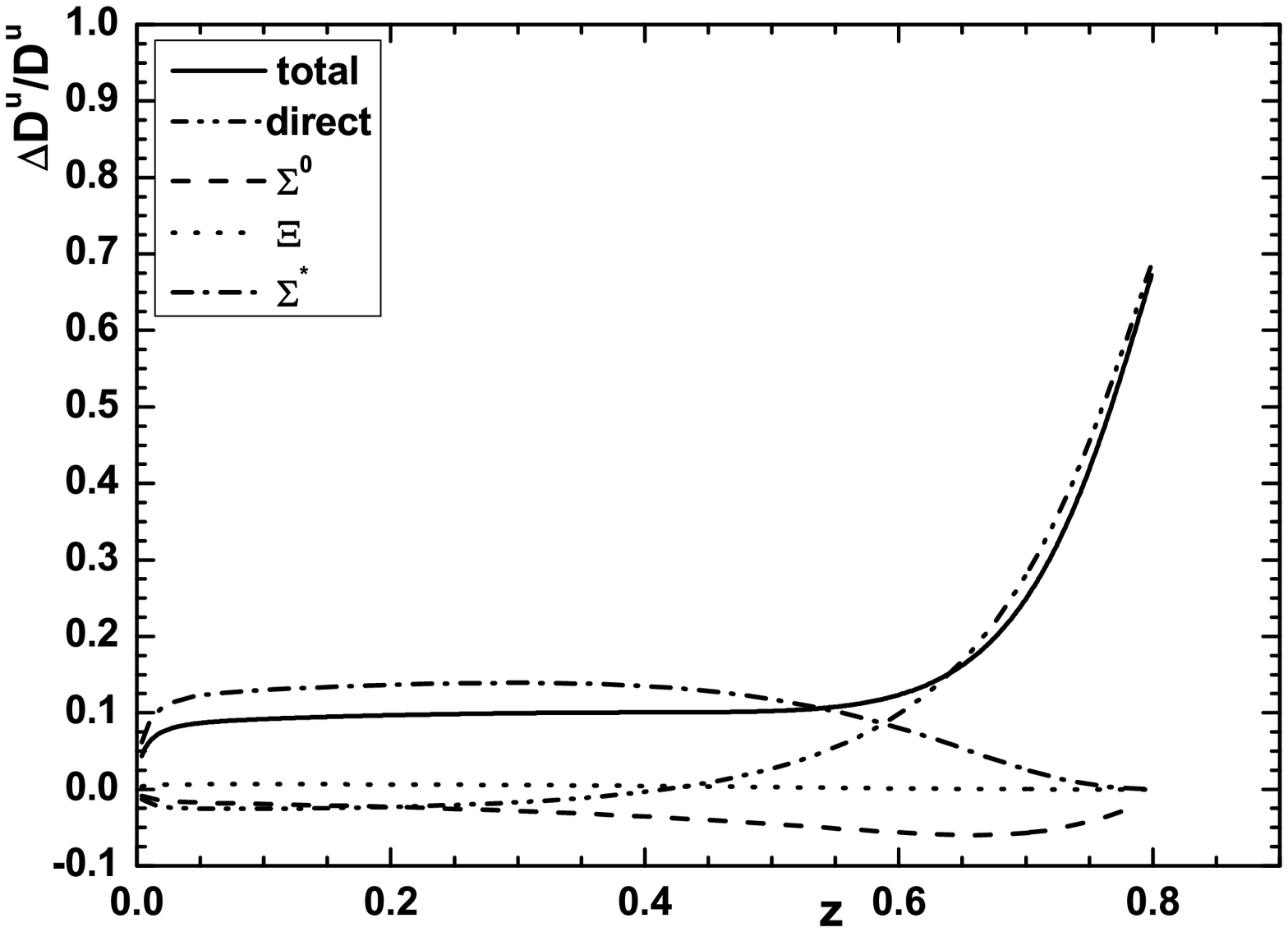}
\includegraphics[width=2.35in]{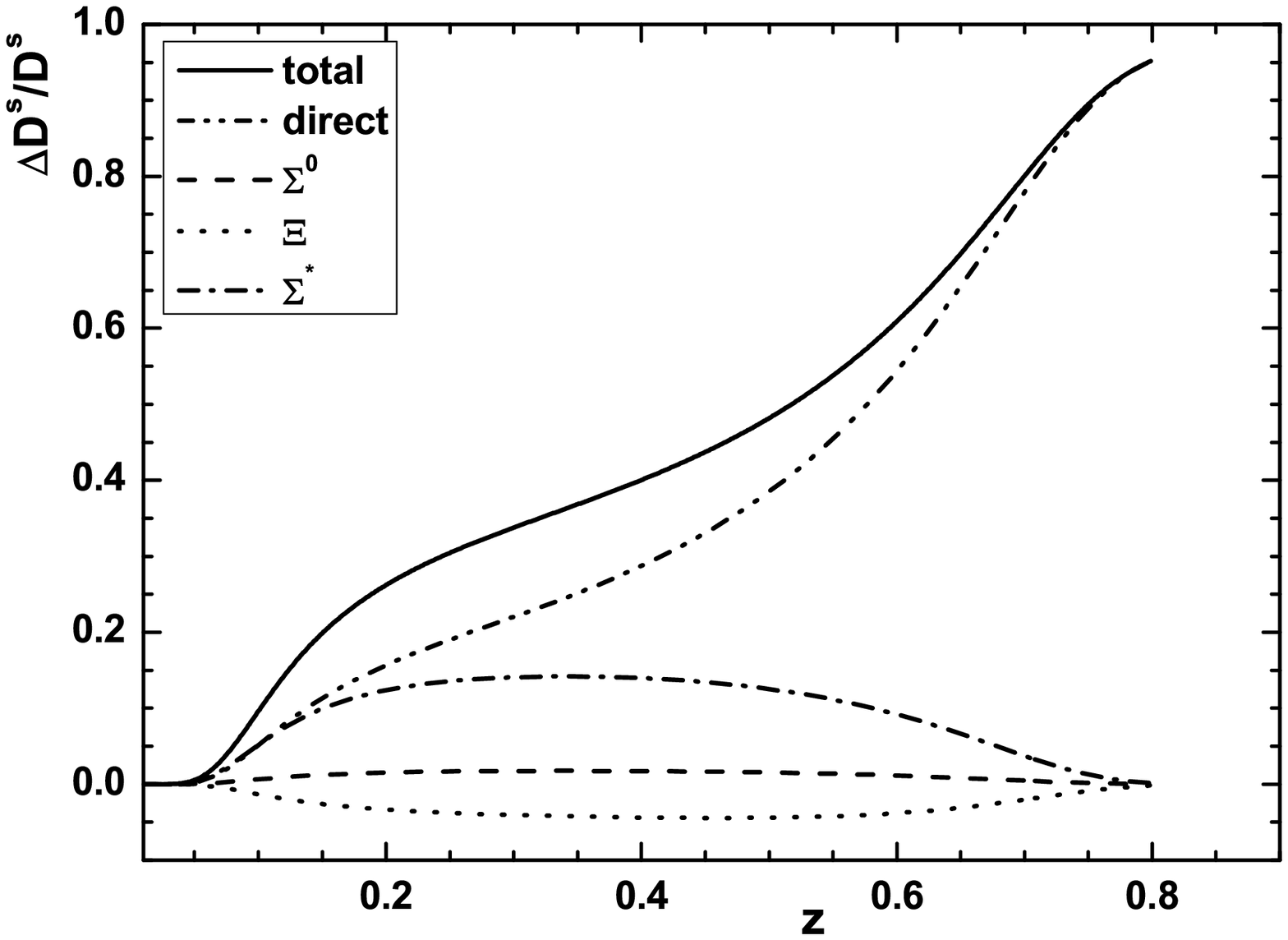}
\caption[*]{\baselineskip 13pt The $z$-dependent ratios of $\Delta D^u(z)/D^u(z)$ and $\Delta D^s(z)/D^s(z)$ with contributions from different channels.}
\label{ratios}
\end{figure}

For the semi-inclusive $eP\rightarrow e \Lambda X$ process, where the electron is longitudinally polarized and the target is unpolarized,
the spin transfer function extracted from the QCD factorization theorem is
\begin{equation}
A^{\Lambda}(z)=\frac{\sum_{q} e_{q}^2xf_{P}^{q}(x,Q^2)
\Delta D^{\Lambda}_{q}(z,Q^2)+(q\rightarrow\overline{q})}
{\sum_{q} e_{q}^2xf_{P}^{q}(x,Q^2)\\
D^{\Lambda}_{q}(z,Q^2)+(q\rightarrow\overline{q})}.
\end{equation}

\begin{figure}%[htb]
 \begin{minipage}[t]{0.5\linewidth}
 %\setcaptionwidth{2.3in}
 %\captionstyle{hang}
  %\centering
  \includegraphics[width=2.8in]{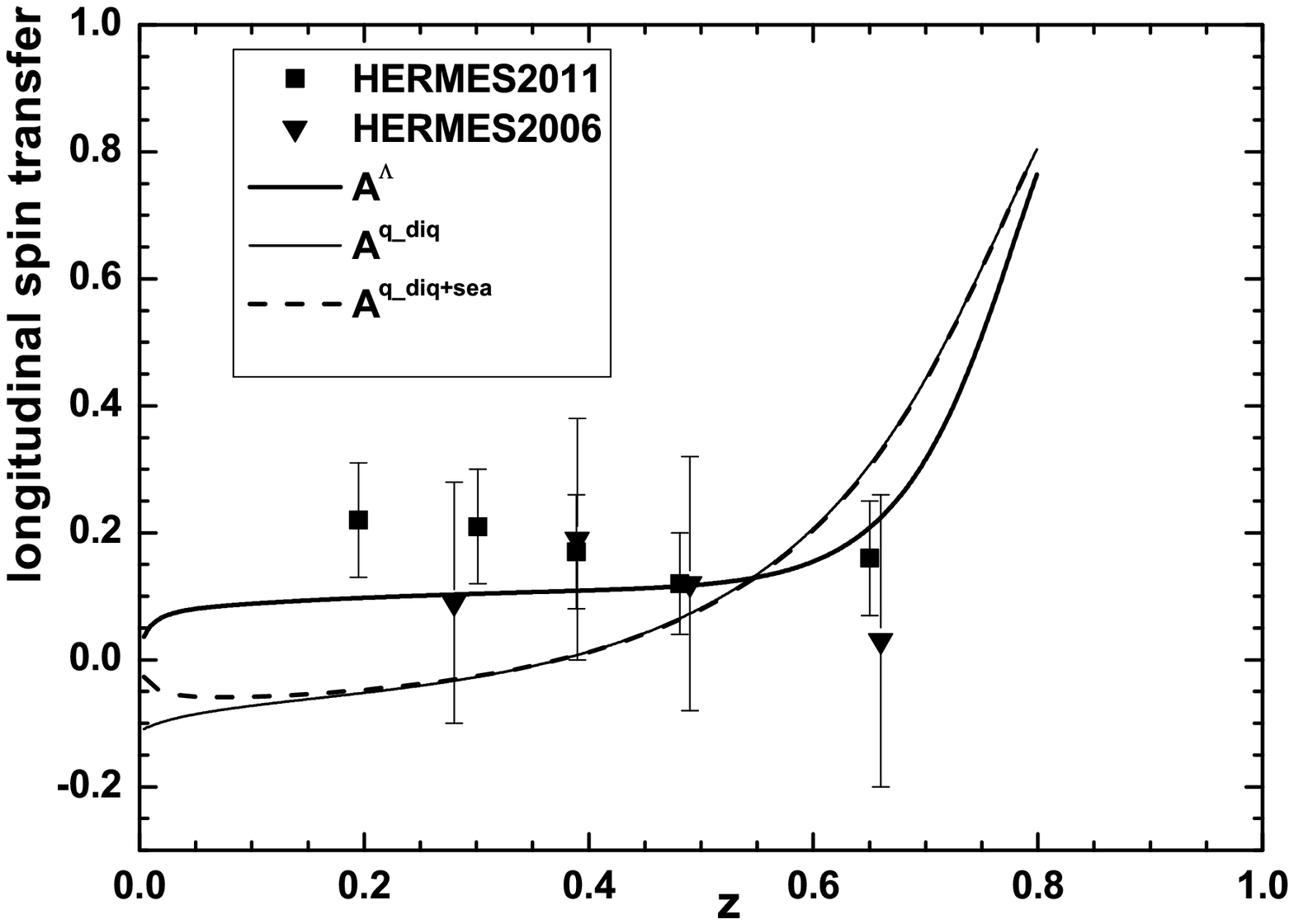}
  \caption[*]{%\baselineskip 10pt
              The results of the
              $z$-dependent longitudinal spin transfer in polarized charged
              lepton DIS process for the $\Lambda$-hyperon.}
              \label{data}
 \end{minipage}%
 \begin{minipage}[t]{0.5\linewidth}
 %\setcaptionwidth{2.3in}
  %\captionstyle{hang}
  %\centering
  \includegraphics[width=2.8in]{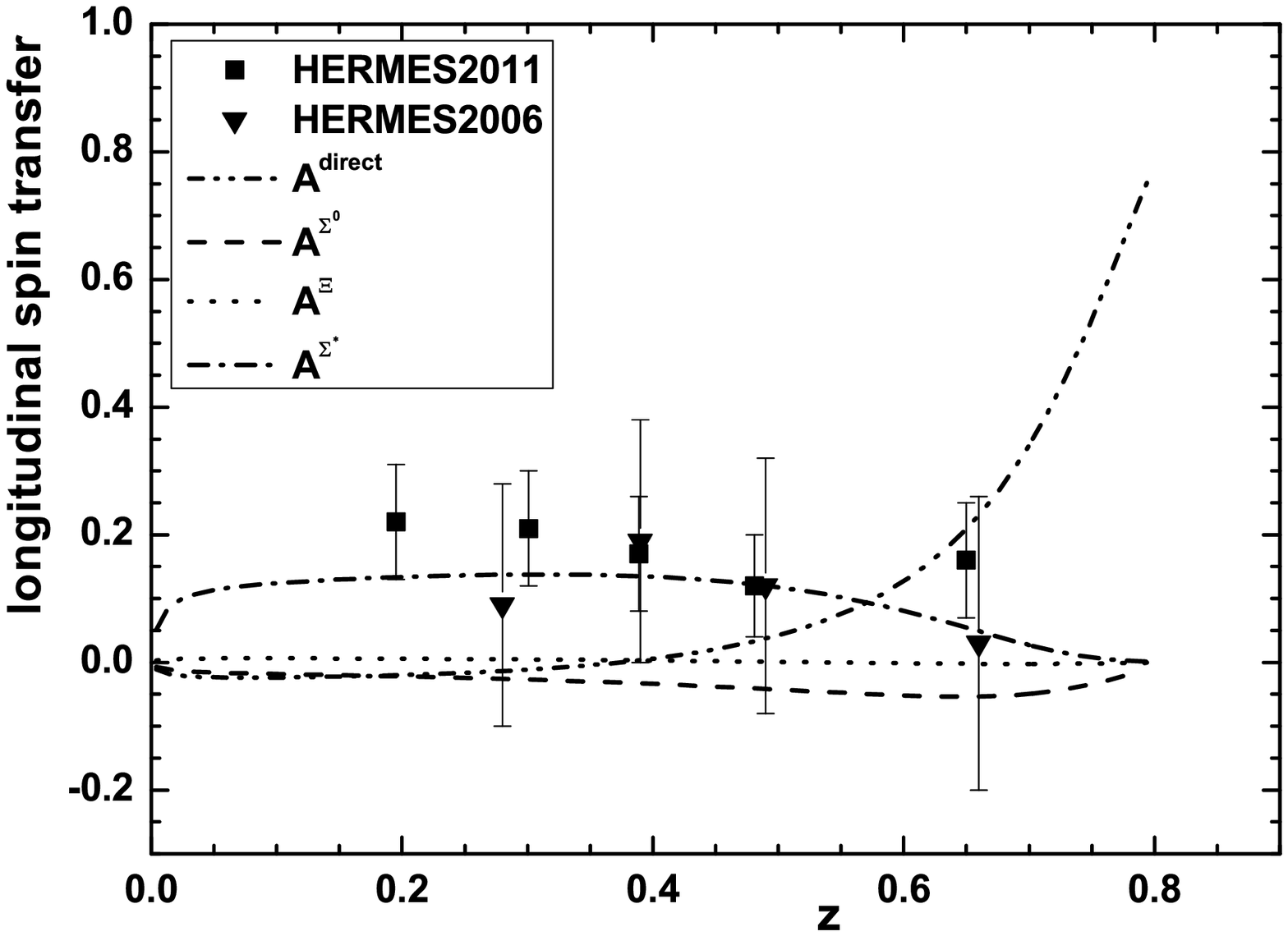}
  \caption[*]{%\baselineskip 10pt
             The results of the
             $z$-dependent longitudinal spin transfer from different channels in polarized charged
             lepton DIS process for the $\Lambda$-hyperon.}
             \label{composition}
 \end{minipage}
\end{figure}

By using our effective $\Lambda$ fragmentation functions and the CTEQ5
parametrization (ctq5l) for the proton, at the point of $x=0.08$, the longitudinal spin transfer distributing
with $z$ is shown in Fig.~\ref{data}. As is shown with the thick solid line in Fig.~\ref{data}, our calculation is well
consistent with the data from the HERMES Collaboration~\cite{Airapetian:2006ee,Belostotski:2011zza}. To
make a comparison, the pure valence quark fragmentation process is calculated using the light-cone SU(6) quark-
diquark model, as shown by the thin solid line. The pure quark and anti-quark fragmentation process is calculated
based on the light-cone SU(6) quark-diquark model and Eq.~(\ref{antiquark}), as shown by the dashed
line. We can see that the anti-quark fragmentation process slightly enhances the spin transfer at the small $z$
region, while the intermediate decay processes greatly improve the spin transfer in the whole $z$ region.

Furthermore, to look at the detailed contribution to this result from different channel, we write the separate longitudinal spin transfer as
\begin{equation}
A^{H_i}(z)=\frac{\sum_{q} e_{q}^2xf_{P}^{q}(x,Q^2)
a_{i}\Delta D^{H_i}_{q}(z,Q^2)\alpha_{H_i\Lambda}+(q\rightarrow\overline{q})}
{\sum_{q} e_{q}^2xf_{P}^{q}(x,Q^2)\sum_{j}
a_{j}D^{H_j}_{q}(z,Q^2)+(q\rightarrow\overline{q})},
\end{equation}
where $H_i$ represents the $\Lambda$ fragmentation contributions from direct fragmentation, the intermediate $\Sigma^0$, $\Xi$, and $\Sigma^{\ast}$ decaying processes respectively.
The results are shown in Fig.~\ref{composition}, in which the $\Sigma^0$ contributes a slightly negative polarization transfer in our plotted region along $z$, but $\Sigma^{\ast}$'s provide a higher positive polarization transfer at low and medium $z$ region, while
the influence from the $\Xi$ is small. We notice that the positive spin transfer at low and medium $z$ region mainly comes from the $\Sigma^{\ast}$ contribution.  This can be easily understood from the following intuitive picture: $\Sigma^{\ast}$ is a spin 3/2 particle composed with three or two positively polarized valence quarks, therefore both the quark to $\Sigma^{\ast}$ fragmentation
and the $\Sigma^{\ast}$ to $\Lambda$ decay process should keep positive spin correlations.

We also examine the longitudinal spin transfer on the $x$ and the Feynman variable $x_F$ dependence. As for the $x_F$,
it can be related to the $x,y,z$ variables through a kinematical transformation.

\begin{figure}%[htb]
 \begin{minipage}[t]{0.5\linewidth}
 %\setcaptionwidth{2.3in}
 %\captionstyle{hang}
  %\centering
    \includegraphics[width=2.8in]{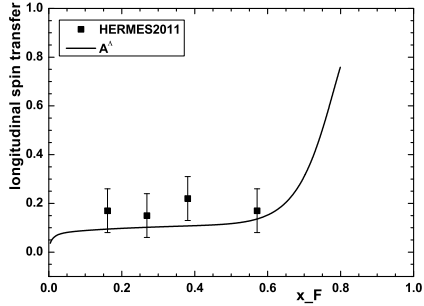}
  \caption[*]{\baselineskip 13pt The result of the
             $x_F$-dependent longitudinal spin transfer in polarized charged
             lepton DIS process for the $\Lambda$-hyperon. The data
              are taken from the HERMES~\cite{Belostotski:2011zza}.}
              \label{longitudinalxF}
 \end{minipage}%
 \begin{minipage}[t]{0.5\linewidth}
 %\setcaptionwidth{2.3in}
  %\captionstyle{hang}
  %\centering
  \includegraphics[width=2.8in]{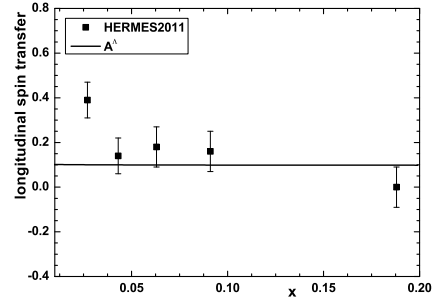}
  \caption[*]{\baselineskip 13pt The result of the
              $x$-dependent longitudinal spin transfer in polarized charged
              lepton DIS process for the $\Lambda$-hyperon. The data
             are taken from the HERMES~\cite{Belostotski:2011zza}.}
               \label{longitudinalx}
 \end{minipage}
\end{figure}

We know in the target rest frame, the four momentum of the proton and the virtual photon are
\begin{equation}
P^{\mu}=(M,0,0,0), ~ q^\mu=(\nu,0,0,-\sqrt{\nu^{2}+Q^2}).
\end{equation}
With a Lorentz transformation, we get the four momentum in the  $\gamma^\ast P$ center of mass frame as
\begin{eqnarray}
P^{\mu}&=&(\gamma M,0,0,-\gamma\beta M), \\ \nonumber%
q^{\mu}&=&(\gamma\nu+\gamma\beta\sqrt{\nu^{2}+Q^2},0,0,-\gamma\sqrt{\nu^{2}+Q^2}-\gamma\beta\nu).\nonumber%
\end{eqnarray}
where $\beta$ is determined by
\begin{equation}
-\gamma\beta M +(-\gamma\sqrt{\nu^{2}+Q^2}-\gamma\beta\nu)=0
\end{equation}
as $\beta = -\frac{\sqrt{\nu^{2}+Q^2}}{M+\nu}$ and $\gamma=\frac{1}{\sqrt{1-\beta^2}}$.
So the invariant mass of the $\gamma^\ast P$ system is
\begin{equation}
W^2=(M+\nu)^2-(\nu^{2}+Q^2).
\end{equation}
The momentum of $P_h$ of the produced hadron $h$ can be parametrized as
\begin{equation}
P_h^\mu\simeq zq^\mu + xzP^\mu + P_{h\bot}^\mu.
\end{equation}
Then the Feynman variable $x_F$ can be obtained as
\begin{eqnarray} \label{eq:xF}
&&x_F=2 P_{hL}/W \nonumber\\
&&~~~= \frac{2z(1-x)\sqrt{\nu^{2}+Q^2}M}{M^2+2M\nu-Q^2} \nonumber\\
&&~~~= z\frac{2(1-x)\sqrt{\frac{Q^4}{4M^{2}x^2}+Q^2}M}{M^2+\frac{Q^2}{x}-Q^2}.
\end{eqnarray}
In our study, as for the $x_F$ bins measured in the HERMES experiment~\cite{Belostotski:2011zza},
we take an average of $Q^2$ as $\bar{Q^2}=4 ~(\mathrm{GeV})^2$
%and for the
%$x$ bins, the average of the $z$ is taken as $z=0.40$, while for the $x_F$ bins, the average of the $x$ is taken as $x=0.09$.
%The results are shown in Fig.~4 and Fig.~5.
and an average of $x$ as $\bar{x}=0.09$, so a collinear transformation of $z$ to $x_F$ can be obtained from Eq.~(\ref{eq:xF}). The calculated result
of the $x_F$-dependent longitudinal spin transfer is then shown in Fig.~\ref{longitudinalxF}. As is shown, the result is in good agreement with the experimental data.

As for the $x$-bins, the calculation is performed at
the average of $\bar{x_F}=0.22$. The result is shown in Fig.~\ref{longitudinalx}. It is known that the cross section of the final hadron produced in the
current-fragmentation region is non-sensitive to the $x$-variable. As is shown in Fig.~\ref{longitudinalx}, our result
is well consistent with this property and the experimental data in the intermediate $x$ region also prove its validity.

The study of the hadronic state produced in the current-fragmentation region of the lepton-lepton annihilation process
can also give information of the quark fragmentations. In this Letter, we consider the process of $e^{+}e^{-}$ annihilation at the $Z$ pole. In this
process, the current $q\overline{q}$ pairs are produced through weak interactions and then fragment into final hadrons.
Though the initial $e^{+}e^{-}$ states are unpolarized, the weak decays can produce polarized current $q\overline{q}$ pairs and these quark pairs
can fragment into polarized $\Lambda$-hyperon. The test of the $\Lambda$ polarization is then sensitive to the
helicity-dependent fragmentations and the specific calculation formula for this process is
\begin{equation}
P_{\Lambda}=-\frac{\sum_{q} A_q \left[{\Delta D_q^h(z)}-
{\Delta D_{\bar q}^h(z)}\right]}{\sum_{q} C_q \left[D_q^h(z)+D_{\bar q}^h(z)\right]},
\label{PL2}
\end{equation}
where the $A_q$ and $C_q$ are parameters as shown in Ref.~\cite{Ma:1999wp}.

\begin{figure}%[htb]
  \begin{minipage}[t]{0.5\linewidth}
 %\setcaptionwidth{2.3in}
 %\captionstyle{hang}
  %\centering
   \includegraphics[width=2.8in]{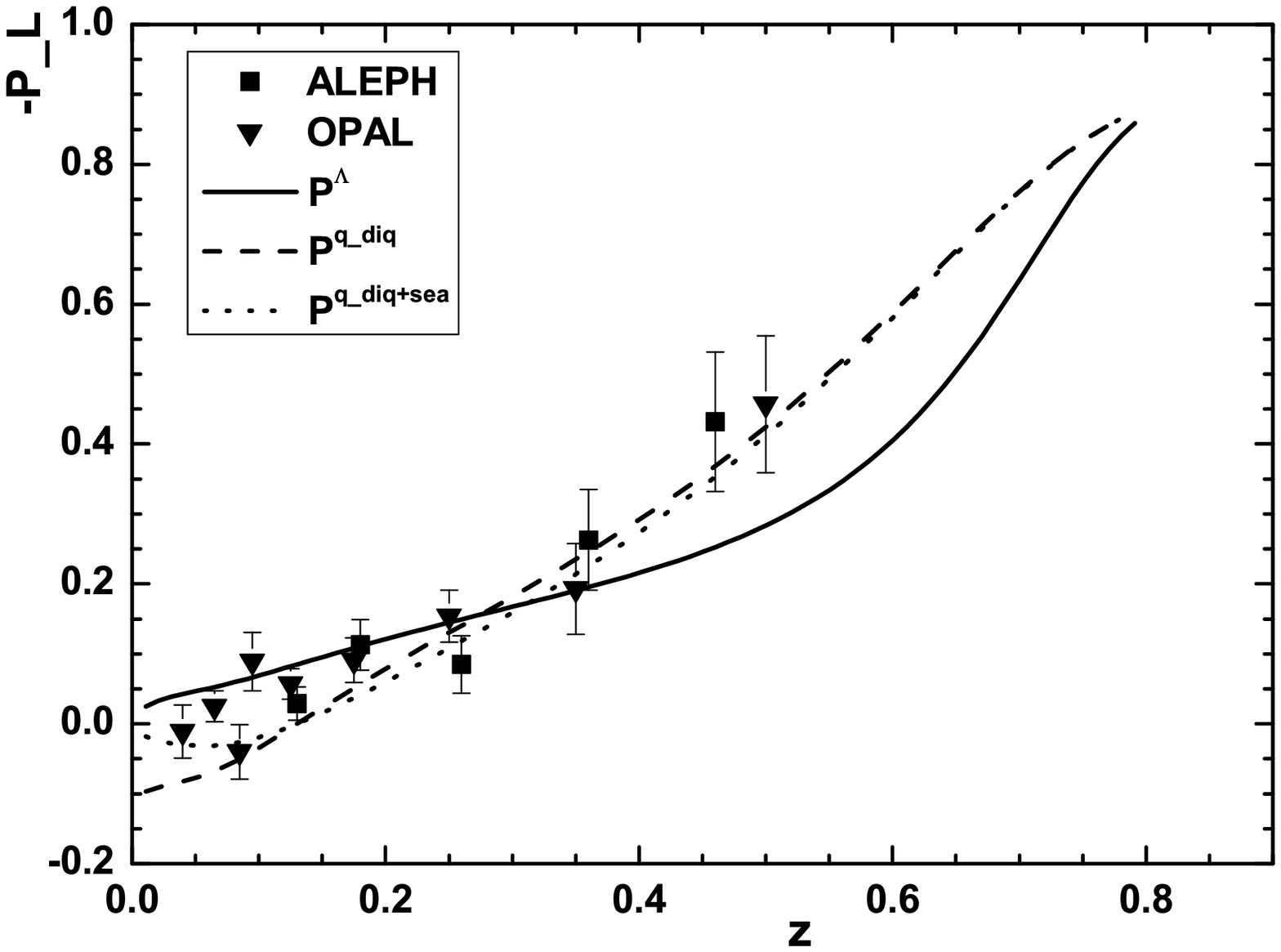}
      \caption[*]{\baselineskip 13pt The results of the
      $z$-dependent longitudinal $\Lambda$ polarization in the
       $e^{+}e^{-}$ annihilation at the $Z$ pole, and the experimental
       data are taken from Refs.~\cite{Buskulic:1996vb,Ackerstaff:1997nh}.}
       \label{z0pole}
\end{minipage}%
 \begin{minipage}[t]{0.5\linewidth}
 %\setcaptionwidth{2.3in}
  %\captionstyle{hang}
  %\centering
  \includegraphics[width=2.8in]{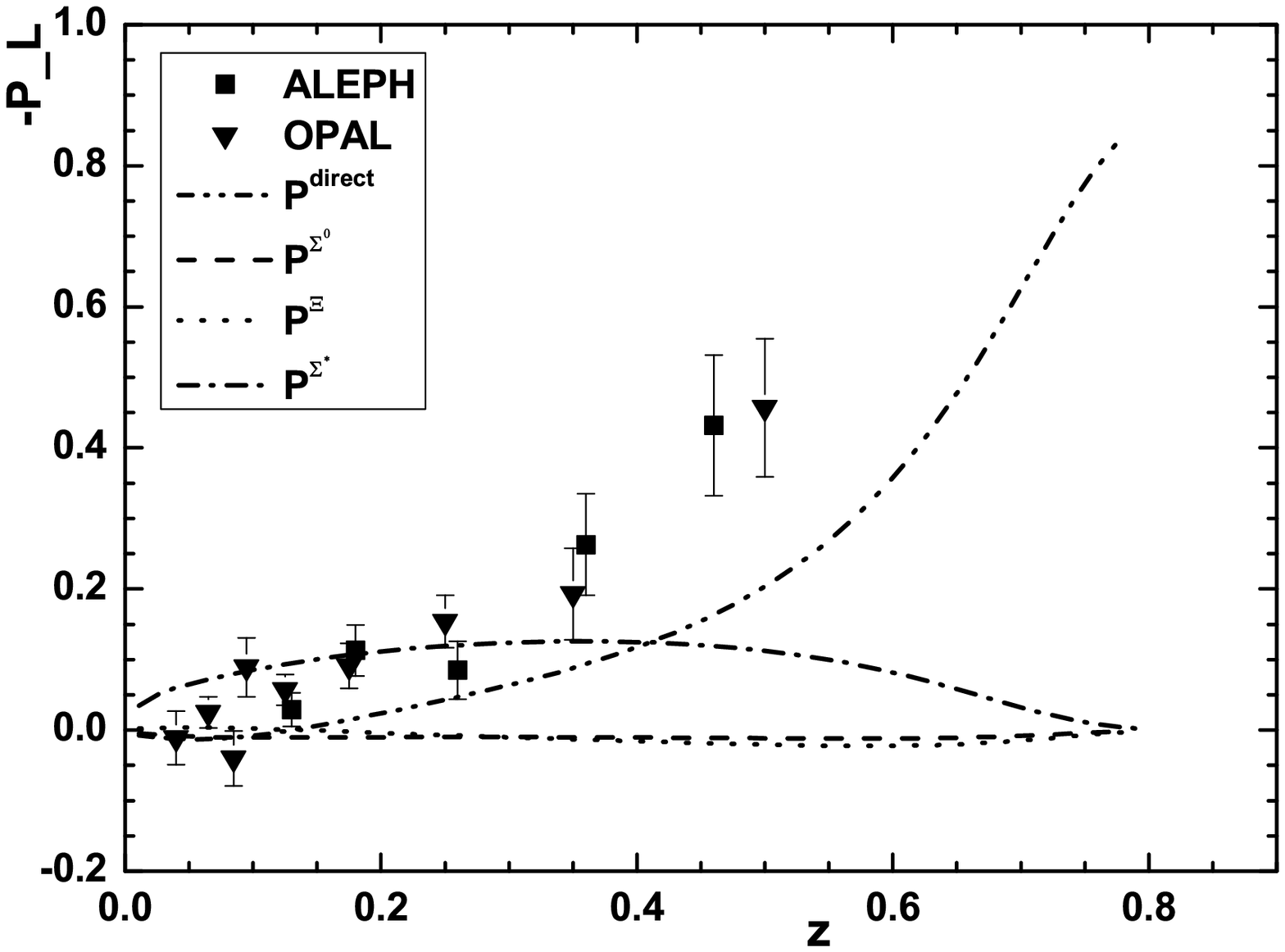}
  \caption[*]{\baselineskip 13pt The results of the
              $z$-dependent longitudinal $\Lambda$ polarization in the
       $e^{+}e^{-}$ annihilation from different channels at the $Z$ pole. The data
             are taken from Refs.~\cite{Buskulic:1996vb,Ackerstaff:1997nh}.}
               \label{separatez0pole}
 \end{minipage}
\end{figure}

Our calculation results are shown in Fig.~\ref{z0pole}. In this figure, the thick solid line is the result of our model, while the thin solid line is
the result from the light-cone SU(6) quark-diquark model with only valence quarks and the dashed line is by including the sea quark content on the
basis of the thin solid line calculation. The result from our effective fragmentation functions agrees with the experimental data.

Detailed contributions to this polarization result from different channels are also considered. The separate contribution is written as
\begin{equation}
P^{H_i}(z)=-\frac{\sum_{q} A_q \left[
a_{i}\Delta D^{H_i}_{q}(z)\alpha_{H_i\Lambda}-(q\rightarrow\overline{q})\right]}
{\sum_{q} C_q \left[\sum_{j}
a_{j}D^{H_j}_{q}(z)+(q\rightarrow\overline{q})\right]},
\end{equation}
where $H_i$ represents the $\Lambda$ fragmentation contributions from direct fragmentation, the intermediate $\Sigma^0$, $\Xi$, and $\Sigma^{\ast}$ decaying processes respectively.
The results are shown in Fig.~\ref{separatez0pole}. As is shown, the main correction to the $\Lambda$ polarization is from the
fragmentation through the $\Sigma^{\ast}$ decay channel, while the $\Sigma^0$ and $\Xi$ contribute slightly.

In summary, we studied the quark to the $\Lambda$ fragmentation properties in the current-fragmentation region
by taking various fragmentation processes into account. These processes include the intermediate decay process
and the anti-quark fragmentation process. By using the light-cone SU(6) quark-diquark model and the Gribov-Lipatov relation,
the effective helicity-dependent
fragmentation functions and unpolarized fragmentation functions are obtained. These effective fragmentation functions are applied to several experimental
processes and the obtained results are in reasonable agreement with experimental data. We thus suggest that
the $\Lambda$-hyperon fragmentation processes are effective to study the structure of the $\Lambda$-hyperon and also
the parton distribution functions of the target particle in the semi-inclusive deep inelastic scattering processes.

This work is partially supported by National Natural
Science Foundation of China (Grants Nos.~11021092, 10975003,
11035003, and 11120101004), by the Research Fund for the
Doctoral Program of Higher Education (China).

\end{document}